\documentclass{cimenics}
\usepackage[ansinew]{inputenc}
\usepackage{pstricks}
\usepackage{amsmath, amssymb}
\usepackage{graphicx}

\title{ELLIPSOIDAL IMPULSE RESPONSES IN FRACTURED MEDIA}

\author{Pedro Contreras (1)\\ Luis Rinc\'on (2)}
\heading{F. Author, S. Author y T. Author}
\address{\noindent \textit{pcontreras@ula.ve}\\\textit{lrincon@ula.ve}\\
(1) Departamento de F\'{\i}sica y Centro de F\'{\i}sica Fundamental, (2) Departamento de Qu\'{\i}mica, Facultad de Ciencias, Universidad de los Andes, M\'erida-Venezuela

\author{Jos\'e Burgos (3)}
\address{\noindent\textit{joseburgos@ula.ve}\\
(3) Universidad Nacional Experimental Sur del Lago, Zulia y Departamento de F\'{\i}sica, ULA}}
\heading{F. Author, S. Author y T. Author}

\abstract{ Multiple vertical fracture sets, combined with horizontal fine layering produce an equivalent medium of orthorhombic or monoclinic symmetry. This is particularly important in fracture reservoir characterization. Fractured reservoirs are azimuthal anisotropic with respect to elastic-wave propagation. In this work we introduce an ellipsoidal approximation for monoclinic media that is able to characterized fractured media near the vertical axis of symmetry. The procedure is basically two-fold. First, we estimate phase velocities near the vertical axis using an expansion of the slowness. Secondly, the phase velocities are used to build the group velocities near the vertical axis. We particularly establish that for monoclinic media ellipsoidal functions in the phase domain correspond to ellipsoidal functions in the group domain. Finally, in order to validate the approximation, the $P$, $S_1$ and $S_2$ ellipsoidal impulse responses are compared for different polar angles with the exact responses obtained by solving numerically the eigenvectors problem from the Christoffel equation. Examples are shown for monoclinic media, and are validated showing results from a previous work for orthorhombic media. The whole procedure is valid for homogeneous media.
}


\begin{document}

\section{INTRODUCCI\'ON}

An orthorhombic model describes a layered medium fractured in two orthogonal directions with azimuthal anisotropic dependence of the group and phase velocities. Wave propagation in orthorhombic media has been extensively studied (see \cite{Musgrave,Helbig} among others) Numerical methods such as finite differences and raytracing using weak anisotropic
\cite{Tsvankin} and elliptical approximation \cite{Contreras} have contributed to the modeling of wavefront in orthorhombic media. Recently Tsvankin \cite{Tsvankin} stressed out the importance of having different approximations to the group wave velocities in order to perform velocity analysis in fracture structures.

A monoclinic model describes two sets of vertical non-corrugated fractures with a horizontal symmetry plane. Monoclinic media have azimuthal anisotropic dependence of the group and phase velocities. There is abundant geological (in situ) \cite{Tsvankin,Schoenberg} evidence of multiple fracture sets, which corroborate the importance of monoclinic models in seismic reservoir characterization; however, velocity analysis, and parameter estimation for monoclinic media is a highly challenging task \cite{Grechka} due to the large number of elastic constants involved.

This paper introduces a mathematical treatment based of the solution of the Christoffer equation in terms of the slowness vector expansion around a vertical axis of symmetry. It is important to choose a coordinate frame where the mathematical description of wave propagation has the simplest form \cite{Grechka}. For instance the equation for the eigenvectors in terms of the slowness and the correspondent eigenvalues are
\begin{displaymath}
F\equiv [G_{il}- \delta_{il}]U_l=0,\;
\qquad \;
F(p_1 \; p_2, \; q = p_3(p_1,p_2)) \equiv  det(c_{ijkl}\; p_j \; p_k - \delta_{il})=0.
\end{displaymath}
$\textbf{U}$ is the polarization vector of a plane wave, $\delta_{il}$ is the Kronecker's delta, and $\hat{\textbf{G}}$ is the symmetric Christoffel matrix, $\textbf{p}$ represents an horizontal slowness vector, and $\textbf{q}$ represents the vertical slowness vector.

For a horizontal reflector the condition is $p_1 = p_2$ $=$ 0; therefore, the zero offset horizontal events in orthorhombic and monoclinic media are obtained by substituting the correspondents values of $\textbf{q}$, and the derivatives $q_{,i}=\frac{\partial q}{\partial p_i}$ $=$ $-\frac{F_i}{F_3}$; where $i$ $=$ $1,2$, and $F_3$ is the derivative of $F$ respect to $q$. The second order derivatives $q_{,ij}$ $=$ $\frac{\partial^2 q}{\partial p_i \partial p_j}$ contain information about the hyperbolic move-out. The derivatives of fourth order $q_{,ijkl}$ $=$ $\frac{\partial^4 q}{\partial p_i \partial p_j \partial p_k \partial p_l }$ contain information about the non-hyperbolic move-out \cite{Tsvankin}. Finally, the slowness can be written as

\begin{equation}\label{ec1}
q(p_1,p_2) = q^0 + q_{,i}|_0 \; p_i + q_{,ij}|_0 \; p_i \; p_j + q_{,ijk}|_0 \; p_i \; p_j \; p_k + q_{,ijkl}|_0 \; p_i \; p_j \; p_k \; p_l.
\end{equation}

From symmetry considerations the derivatives of first $q_{,i}|_0$, and third order $q_{,ijk}|_0$, are equal to zero for both types of symmetry. The derivatives of fourth order are taken to be zero since non-hyperbolic moveout is neglected. Eq. (\ref{ec1}) yields suitable expressions for the vertical slowness $\textbf{q}$ in the orthorhombic and the monoclinic cases for the three propagation modes; namely $P$, $S_1$ and $S_2$.

In voigt's notation the elastic tensor for the monoclinic structure has the form \cite{Musgrave,Helbig,Grechka}
\begin{equation}
C_{MN} = \left(
              \begin{array}{cccccc}
                 C_{11} & C_{12} & C_{13} & 0 & 0 & C_{16} \\
                 C_{21} & C_{22} & C_{23} & 0 & 0 & C_{26} \\
                 C_{31} & C_{32} & C_{33} & 0 & 0 & C_{36} \\
                 0 & 0 & 0 & C_{44} & C_{45} & 0 \\
                 0 & 0 & 0 & C_{45} & C_{55} & 0 \\
                 C_{16} & C_{26} & C_{36} & 0 & 0 & C_{66} \\
               \end{array}
               \right)\nonumber\\
\end{equation}

Orthorhombic symmetry is described by nine elastic constants (with $C_{16}$ $=$ $C_{26}$ $=$ $C_{36}$ $=$ $C_{45}$ $=$ 0). It has two vertical planes of symmetry, $[XZ]$, and $[YZ]$. There is not general solution for the Christoffel equation for orthorhombic symmetry; however, in both symmetry planes are possible analytical solutions. For other angles, a numerical solution is given for each mode by the roots of the characteristical polynomial \cite{Helbig,Contreras}.

Monoclinic symmetry is described by thirteen elastic constants, but if $C_{45}$ becomes zero, then the matrix becomes diagonal for vertical propagation and the horizontal axes coincide with the polarization directions of the vertical traveling shear wave \cite{Musgrave,Grechka}. Therefore, monoclinic symmetry is described by twelve elastic constants. There is not general solution for the Christoffel equation for monoclinic symmetry, the solution always requires a numerical treatment for each propagation mode. The eigenvalue method through the characteristical polynomial gives the phase velocities and the eigenvector method for \textbf{U} yields the group velocities \cite{Helbig,Grechka,Zhou}.

\section{ELLIPSOIDAL PHASE VELOCITIES (WAVEFRONTS) NEAR THE VERTICAL AXIS}
Eq. (\ref{ec1}) yields the ellipsoidal phase velocities (wavefronts) near the vertical axis \cite{Contreras,Grechka}; however, the correspondent derivation is long, and we refer for a detailed description of the method to look at appendix of reference \cite{Grechka}.
A propagation shear mode separation always occurs in fracture media, where two shear velocities always exists. The substitution of $q_{,ij}|_0$ $=$ $0$ in Eq.(\ref{ec1}) using a slowness vector $\vec{p}$ $=$ $(p_1,p_2,q)$, and a unitary phase vector $\vec{n}$ $=$
$(\sin\theta_{i1}\cos\theta_{i2},\sin\theta_{i1}\sin\theta_{i2},\cos\theta_{i1})$ yields the square of the phase velocities $W_i$ $=$ $V^2_i \; \rho$ (where $\rho$ is the density). The calculation for the monoclinic media near the vertical axis gives the following expression

\begin{equation}\label{ec2}
W^i = W^{i,z} \cos\theta^2_{i1} + \sin\theta^2_{i1}
\Big( W^{i,nmo}_{11}\cos\theta^2_{i2} + 2W^{i,nmo}_{12}\sin\theta_{i2}\cos\theta_{i2} + W^{i,nmo}_{22}\sin\theta^2_{i2}.\Big)
\end{equation}
"Eq. (\ref{ec2})" represents an ellipsoid in the phase domain for each elastic mode $i$ $=$ $P$, $S_1$ and $S_2$. For waves traveling in the vertical direction: $W^{P,z}$ $=$ $\frac{1}{\sqrt{C_{33}}}$, $W^{S1,z}$ $=$ $\frac{1}{\sqrt{C_{55}}}$, and $W^{S2,z}$ $=$ $\frac{1}{\sqrt{C_{44}}}$.

For the longitudinal mode ($P$), the square of the horizontal NMO velocities for the vertical symmetry planes $[11]$ and $[22]$ are
\begin{eqnarray*}
&& W^{P,nmo}_{[11]} = \frac{(C_{33}- C_{44})(C_{13}^2+2C_{13}C_{55}+ C_{33}C_{55})+ C_{36}^2 (C_{33}-C_{55})}{(C_{33}- C_{44})(C_{33}-C_{55})},\nonumber\\
&& W^{P,nmo}_{[22]} = \frac{(C_{33}- C_{55})(C_{23}^2+2C_{23}C_{44}+ C_{33}C_{44})+ C_{36}^2 (C_{33}-C_{44})}{(C_{33}- C_{44})(C_{33}-C_{55})},\nonumber\\
\end{eqnarray*}
outside the symmetry planes the NMO velocity is given by the $[12]$ term
\begin{equation*}\label{ec2}
W^{P,nmo}_{[12]} = C_{36} \; \frac{\Big[(2C_{55}+ C_{13})- C_{33}(C_{13}+C_{23}+C_{44}+C_{55})\Big]}{(C_{33}-C_{44})(C_{33}-C_{55})}.\nonumber\\
\end{equation*}
For the shear mode ($S_1$), the square of the NMO velocities are:
\begin{displaymath}
W^{S1,nmo}_{[11]}=C_{11}+\frac{[C_{13}+ C_{55}]^2}{C_{33}-C_{55}},\nonumber\\
\qquad
W^{S1,nmo}_{[12]}=C_{16}+\frac{C_{36}[C_{13}+ C_{55}]}{C_{33}-C_{55}},\nonumber\\
\qquad
W^{S1,nmo}_{[22]}=C_{66}+\frac{C^2_{36}}{C_{33}-C_{55}}.
\end{displaymath}
Finally, for the shear mode ($S_2$) the square of the NMO velocities are:
\begin{displaymath}
W^{S2,nmo}_{[11]}=C_{66}+\frac{C^2_{36}}{C_{33}-C_{44}},\nonumber\\
\qquad
W^{S2,nmo}_{[12]}=C_{26}+\frac{C_{36}[C_{23}+ C_{44}]}{C_{33}-C_{44}},\nonumber\\
\qquad
W^{S2,nmo}_{[22]}=C_{22}+\frac{[C_{23}+ C_{44}]^2}{C_{33}-C_{44}}.
\end{displaymath}

\section{ELLIPSOIDAL GROUP VELOCITIES (IMPULSE RESPONSES) NEAR THE VERTICAL AXES}
In order to generate expressions in monoclinic media for group velocities within the ellipsoidal approximation at small group polar angles $\phi_{1i}$, and arbitrary azimuthal group angles $\phi_{2i}$, we rely on a transformation using in Ref.\cite{Byum} for VTI media, and in Ref.\cite{Contreras} for orthorhombic media. These transformations are proposed and used here as follows:
\begin{displaymath}
\theta_{1i} \rightarrow \phi_{1i},
\qquad
\theta_{2i} \rightarrow \phi_{2i},
\qquad
W_{i,phase,monoclinic} \rightarrow W^{-1,i}_{group,monoclinic}.
\end{displaymath}
$\phi_{1i}$ are the polar, and $\phi_{2i}$ are azimuthal group velocities angles. Then, ellipsoidal group velocities near the vertical axis are ellipsoids in the group domain.
The inverse of the square of the group velocity (impulse response) obtained from "Eq. (\ref{ec2})" for each propagation mode ($i$ $=$ $P$, $S_1$, $S_2$) is
\begin{equation}\label{ec3}
W^{-1,i}_{group} = \frac{\cos\phi^2_{i1}}{W^{i,z}} + \sin\phi^2_{i1}
\Big( \frac{\cos\phi^2_{i2}}{W^{i,nmo}_{[11]}} + \frac{2\sin\phi_{i2}\cos\phi_{i2} }{W^{i,nmo}_{[12]}}+ \frac{\sin\phi^2_{i2}}{W^{i,nmo}_{[22]}}\Big),
\end{equation}
where $W^i_{group}$ $=$ $ \rho V^{2,i}_{group}$.

\section{NUMERICAL EXAMPLES}

The validation of our approximation is performed on the cracked Greenhorn shale case of orthorhombic symmetry, but by aggregating a non perpendicular fracture set, according to
the Muir-Schoenberg theory \cite{Schoenberg,Dellinger}, the elastic matrix then, becomes monoclinic. The values of the elastic constants are: $C_{11}= 336.6$, $C_{12}= 117.3$, $C_{13}=103.3$, $C_{22}=310.0$, $C_{23}=92.3$, $C_{33}=223.9$, $C_{44}=49.1$, $C_{55}=54.0$, $C_{66}=94.6$, $C_{16}=30.0$, $C_{26}=30.0$ and $C_{36}=10.0$. As in Ref. \cite{Dellinger} we have chosen $C_{ij}$ to be dimensionless, and the density $\rho$ has been set to unity

The exact group velocity (impulse response) can be found numerically with the expression \cite{Helbig,Zhou}
\begin{equation}\label{ec4}
V^i_{group} = C_{ijkl} \alpha_j \; \alpha_k \; \beta_l,
\end{equation}
where $C_{[ijkl]}$ are the elastic constants in the usual notation, $\alpha_j$ represents the eigenvectors that correspondent to the unit polarization vectors, $\beta_l$ correspond to the normal direction of the wavefront, Eq.(\ref{ec4}) will be numerically solved in order to compare with Eq.(\ref{ec3})

\begin{figure}[t]
\begin{center}
\includegraphics[width = 5.0 in, height= 2.5 in]{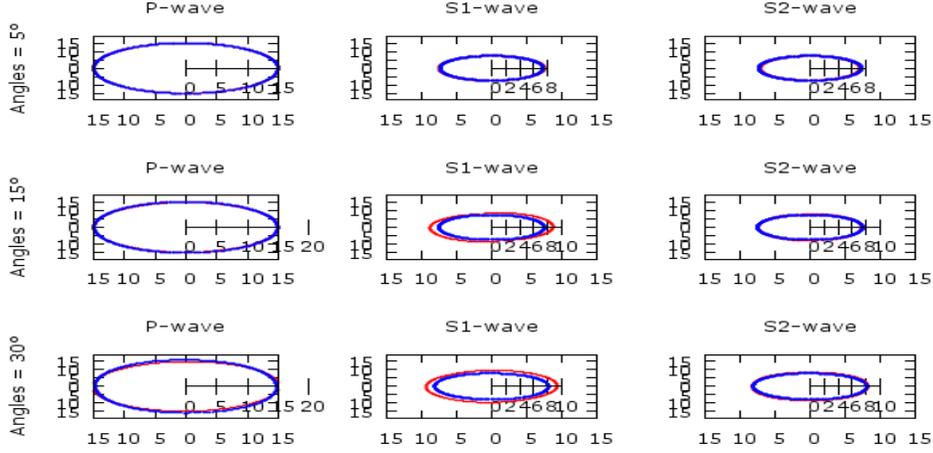}
\caption{\label{FS1}Azimuthal view of each wave-mode impulse response for different polar angles $\phi_{1i}$ in the monoclinic case. Red color represents the exact solutions from Eq.(\ref{ec4}), blue color represents the ellipsoidal approximation of Eq. (\ref{ec3}).}
\label{mono}
\end{center}
\end{figure}

\begin{figure}[t]
\begin{center}
\includegraphics[width = 4.0 in, height= 2.0 in]{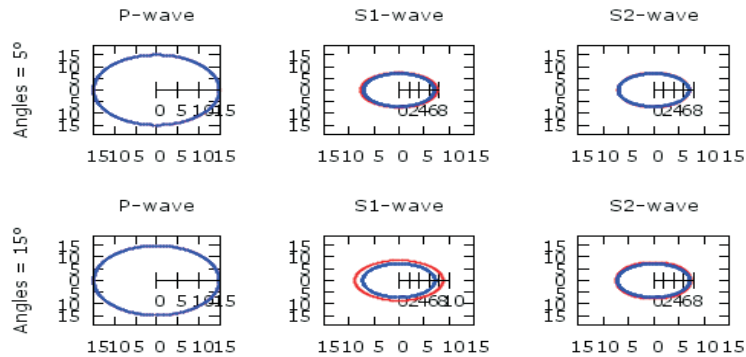}
\caption{\label{FS2}Azimuthal view of each wave-mode impulse response for different polar angles $\phi_{1i}$ in the orthorhombic case reproduced according to \cite{Contreras} Red color represents the exact solutions from Eq.(\ref{ec4}), blue color represents the ellipsoidal approximation of Eq. (\ref{ec3}).}
\label{ortho}
\end{center}
\end{figure}

Figure (\ref{mono}) shows horizontal slices of the 3-D impulse responses for each wave propagation mode at different polar angles in the monoclinic case. It can be observed that the exact group velocity (red color) and the ellipsoidal group velocities (blue color) are almost the same for polar angles with small vertical aperture. As the polar angle $\phi_{1i}$ increases their separation also increases, and the ellipsoidal approximation begins to deteriorate. The maximum approximation error is reached at a polar angle of $90^{0}$ (not shown here).
Figure (\ref{ortho}) shows horizontal slices of the 3-D impulse responses for each propagation mode at different polar angles $\phi_{1i}$ in the orthorhombic case. It can be observed that exact (red color) and ellipsoidal (blue color) group velocities are almost the same at the $[XY]$ symmetry plane for $\phi_{1i}$ = $5^0$, but at the polar angle $\phi_{1i}$ of $15^{0}$ the elliptical approximation of the shear modes begin to deteriorate, as it was shown in Ref. \cite{Contreras}; hence, the horizontal velocity is not well reproduce for large $\phi_{1i}$.


\section{CONCLUSIONS}
The impulse responses of all different modes of wave propagation in monoclinic media near the vertical axis are ellipsoids Figure (\ref{mono}). This implies that the ideas depicted in \cite{Contreras,Byum} are applicable to monoclinic media as well.
On the other hand, it is shown that the horizontal Normal Move-Out velocities for horizontal reflectors are ellipses for monoclinic media in agreement with the results of \cite{Grechka}.
It is found that the off diagonal NMO velocities are controlled in monoclinic media for three elastic parameters: the longitudinal mode $W^{P,nmo}_{[12]}$ is controlled by $C_{36}$, the transversal $S_1$ mode $W^{S1,nmo}_{[12]}$ is controlled by $C_{16}$,
and the transversal $S_2$ mode $W^{S2,nmo}_{[12]}$ is controlled by $C_{26}$. By making these three elastic constants to be zero, the orthorhombic case of Ref.\cite{Contreras} is reproduced Figure (\ref{ortho}).

Therefore, we particularly establish that for monoclinic media ellipsoidal functions in the phase domain correspond to ellipsoidal functions in the group domain.
The ellipsoidal approximation is therefore a simple but a powerful device to reproduce elastic group velocities in fractured media accurately near the vertical axis.

\section*{\textit{Acknowledgements}}

We thank Dr. V. Grechka  for stimulating discussions . We also
acknowledge discussions with Profesores D. Guti\'errez, R. Almeida and L. Seijas
from the University of los Andes. This research was supported by the Grant
CDCHTA-ULA C-1851-13-05-B.

\end{document}